# Dispersal and dive patterns in gravid leatherback turtles during the nesting season in French Guiana

Sabrina Fossette[1,2], Jean-Yves Georges[1*], Hideji Tanaka[3,4], Yan Ropert-Coudert[5], Sandra Ferraroli[1], Nobuaki Arai[3], Katsufumi Sato[5], Yasuhiko Naito[5] and Yvon Le Maho[1]

[1] Centre National de la Recherche Scientifique, Institut Pluridisciplinaire Hubert Curien, UMR 7178 CNRS-Université Louis Pasteur, Département d'Ecologie, Physiologie et Ethologie, 23 rue Becquerel, 67087 Strasbourg, France

[2] Université Louis Pasteur, 4 rue Blaise Pascal, 67070 Strasbourg, France

[3] Department of Social Informatics, Graduate School of informatics, Kyoto University, 606-8501 Kyoto, Japan

[4] COE for Neo-Science of Natural History, Graduate School of Fisheries Sciences, Hokkaido University, 041-9611 Hakodate, Japan

[5] National Institute of Polar Research, 1-9-10 Kaga, Itabashi-Ku, Tokyo 173-8515, Japan

\* Corresponding author

Jean-Yves Georges

Phone: +33 388 106 947

Fax: +33 388 106 906

Email: jean-yves.georges@c-strasbourg.fr

# ABSTRACT

We present the first combined analysis of diving behaviour and dispersal patterns in gravid leatherback turtles during 3 consecutive nesting seasons in French Guiana. In total 23 turtles were fitted with Argos satellite transmitters and 16 individuals (including 6 concurrently satellite-tracked) were equipped with an electronic time-depth recorder for single inter-nesting intervals, i.e. between two consecutive ovi-positions. The leatherbacks dispersed over the continental shelf, ranging from the coastal zone to the shelf break and moved over 546.2 ± 154.1 km (mean ± SD) in waters of French Guiana and neighbouring Surinam. They mostly performed shallow (9.4 ± 9.2 m) and short (4.4 ± 3.4 min) dives with a slight diurnal pattern. They dived deeper as they moved away from the coast suggesting that they were predominantly following the seabed. Inter-nesting intervals could be divided into two phases: during the first 75% of the time turtles spent at sea, they dived on average 47 min $h^{-1}$ before showing a lower and more variable diving effort as they came back to the shore. The extended movements of leatherbacks and the fine analysis of dive shapes suggest that in French Guiana leatherbacks may feed during the inter-nesting interval, probably to compensate for the energy costs associated with reproduction. This results in this endangered species being exposed to high risks of interactions with local fisheries throughout the continental shelf.

# INTRODUCTION

Understanding how wild animals face trade-offs between survival and reproduction is important in species with high reproductive effort, particularly in critically endangered species where adult mortality may be high enough to result in extinction. In sea turtles, reproduction takes place over 2 months during which females lay 1 to 14 clutches of 50 to 130 eggs each, depending on the species (Miller 1997). Accordingly, sea turtles (IUCN 2004) have high reproductive energy costs. Among sea turtles, critically-endangered leatherback turtles (*Dermochelys coriacea*) show the highest reproductive output (Miller 1997) as they may lay the equivalent of 20% of their body mass (Georges et al. unpublished data) and supposedly do not feed over the nesting season (Miller 1997, Rivalan et al. 2005). Consequently, during the inter-nesting intervals, leatherbacks might minimise energy expenditure to maximize the amount of energy allocated to ovi-position and egg production (Reina et al. 2005, Wallace et al. 2005). This seems to be the case in the gravid leatherbacks nesting in Pacific coasts of Costa Rica, which have been reported to move very slowly near the seabed (Reina et al. 2005) and display an extremely low metabolic rate (Wallace et al. 2005) during the nesting season.

In contrast, the gravid leatherbacks nesting on both sides of the Atlantic cover great distances at sea (Georges et al. in press) and swim at high speed during inter-nesting intervals (Eckert et al. 1989, Eckert 2002). Atlantic leatherbacks nesting in the Caribbean perform nocturnal shallow, and diurnal deep, dives (Eckert et al. 1989) consistent with the vertical migration of their main prey (i.e. gelatinous plankton, Hays 2003), auguring for nocturnal foraging activity (Myers & Hays in press). In other words, between two consecutive nesting events, leatherback turtles may compensate for high reproductive costs either by reducing their activity, or by feeding, as suggested for the Pacific and Atlantic populations, respectively.

The diving behaviour of marine animals has been studied in several different ways, such as by analysing depth profiles concurrently with, by example, swim speed (e.g. Ropert-Coudert et al. 2000, Eckert 2002), three-dimensional compass data (e.g. Mitani et al. 2003) or underwater video (e.g. Reina et al. 2005, Watanabe et al. 2006). Concurrent study of diving and dispersal behaviour (Georges et al. 1997) provides important information regarding the areas where particular behaviours occurred, and defines the oceanic zones where individuals tend to congregate. Such information is crucially needed for protected species whose distribution overlaps with areas of natural and/or anthropogenic threats (e.g. Georges et al. in press).

Here we present the first combined analysis of diving and dispersal patterns of the critically-endangered leatherback turtle during their inter-nesting intervals over 3 consecutive nesting seasons in French Guiana. Following the extended dispersal recently reported in this population (Georges et al. in press), we predict that leatherbacks in French Guiana do not reduce their activity as suggested in the Pacific ocean (Reina et al. 2005), but may rather dive consistently and display feeding activity, as suggested in the Caribbean Sea (Eckert et al. 1989, Myers & Hays in press). In addition, since leatherbacks face lethal interactions with industrial fisheries in French Guiana while dispersing widely over the continental shelf (Delamare 2005, Georges et al. in press), our study aims to identify those areas and depths where these interactions are more likely to occur.

## MATERIALS AND METHODS

The study was carried out during the nesting seasons 2001-2003 at Awala-Yalimapo beach (5.7°N – 53.9°W), French Guiana, on the border with Surinam, South America (**Figure 1**).

**Horizontal movements.** During the study period, a total of 23 females was equipped with satellite platform transmitter terminals (PTTs, Kiwisat 101 AA-cell, Sirtrack, New-Zealand, weight 150g, cross section: 4cm²). PTTs were held in place on the carapace using a customised harness attached during ovi-position in 2001 (see Eckert et al. 1996) and were directly fixed on the central ridge of the turtle's carapace in 2002 and 2003 (see Southwood et al. 1999). Harnesses were automatically released from the animals after several months during post-nesting migrations due to a corrodible link in the attachment system (Eckert et al. 1996). PTTs directly fixed on the carapace were removed as soon as turtles came back to the nesting beach after at least one inter-nesting interval.

At-sea movements were reconstructed using the Argos system (www.cls.fr). Each Argos location was provided with a class of accuracy, with classes 1, 2 and 3 having nominal standard deviations around the true position of 1000 m, 350 m and 150 m, respectively, whereas location classes A, B and 0 have no designed accuracy. We analysed all locations of all accuracies, excluding locations that were on land, locations separated by less than one hour and locations that implied travel rates > 10km/h (Eckert 2002, Gaspar et al. 2006) by filtering out the least accurate locations.

For individual turtles tracked during more than one inter-nesting interval, we only considered the first track to avoid pseudo-replication. For each individual track, we calculated the time spent at sea, the overall distance travelled and the distance to the furthest point from the beach (dispersal range). Each track was divided into phases according to the distance the turtles moved relative to the beach. Outbound/inbound phases corresponded to the period when turtles moved away/back from/to the beach, respectively. For some individuals an intermediate phase between outbound and inbound phases was identified, when the distance to the beach remained close to its maximum value.

**Vertical movements.** Sixteen females fitted with a PTT were concurrently equipped with an electronic Time-Depth Recorder (TDR, Little Leonardo, Japan, weight: 54g, cross section: 3.5cm², length: 116mm), directly fixed on the central ridge of the turtle's carapace for a single inter-nesting interval. Each logger included a pressure sensor measuring depth (range: 0-100 m, ±0.05m) every second. Following recapture, loggers were removed and data downloaded into a laptop computer. Data were analysed using IGOR Pro software (WaveMetrics Inc., Oregon, USA). All dives > 0.5 m and < 2 m (n = 48220 dives) occurred exclusively during the first 1.5h and the last 6h of each individual inter-nesting interval, probably reflecting travelling from/to the nesting beach, and were excluded from the analysis to allow us to focus on other diving patterns. We recorded the start and end time of each dive, the maximum depth reached, the duration of descent/bottom/ascent phases, the rates of descent and ascent and the duration of the preceding and the subsequent post-dive surface interval. The bottom phase was defined as the period during which depth was deeper than 90% of the maximum depth of a given dive. For each bottom phase, we calculated the depth amplitude and number of rapid, up-and-down undulations observed in the depth profile (hereafter termed as wiggles, see Wilson 1995 and Houghton et al. 2002).

In order to classify dive profiles, a Principal Component Analysis (PCA) was performed on all 20607 dives > 2 m considering the above-mentioned parameters. As the total number of dives recorded varied among individuals (from 679 to 3539 dives), relationships between dive parameters were performed considering a random sample of 600 dives per individual. This avoided pseudo-replication while taking inter-individual variability into account (Cherel et al. 1999). Diurnal patterns in terms of number of dives, dive depth and dive duration were investigated considering the nautical definition of dawn and dusk (i.e. when the sun was 12° below the horizon at Awala-Yalimapo beach; http://aa.usno.navy.mil/).

Diving effort was analysed by considering hourly dive frequency and time spent diving per hour. Changes in hourly diving effort were investigated throughout the inter-nesting interval in 12-hour increments centred on midnight and midday, excluding 12-hours blocks that were incomplete (i.e. for the first and last hours of the trip). Statistical analyses were carried out using Minitab statistical software. Values are given as means ± SD, differences being considered as statistically significant when $P < 0.05$.

All turtles were measured during ovi-position using a flexible measuring tape (±0.5cm) following Georges & Fossette (2006). Standard curvilinear carapace length was measured on the midline of the shell, from the nape notch of the carapace to the end of the caudal peduncle. Curvilinear carapace width was measured at the level of maximum width back of the fore-flippers (Georges & Fossette 2006).

## RESULTS

Among the 23 females equipped with ARGOS transmitters, we obtained data from 11 individuals (7 in 2001, 2 in 2002 and 2 in 2003) over at least one complete inter-nesting interval (**Table 1**). Among the 16 individuals that were fitted with TDRs, 10 were successfully recaptured with their tag, within which 7 (3 in 2001, 2 in 2002 and 2 in 2003; **Table 2**) were monitored over one complete inter-nesting interval. Four individuals were concurrently monitored with Argos and TDR devices over one complete inter-nesting interval.

**Horizontal movements**

Most locations obtained from Argos were of poor quality, with locations 3, 2, 1, 0, and below contributing to 4, 8, 14, 13 and 60% of the total number of locations received, respectively. The 11 turtles spent on average 10.2 ± 0.9 days (range: 8.3-11.8 days) at sea between two consecutive nesting events (**Table 1**). Turtles dispersed within a short range from the coast (90.4 ± 47.7 km, range: 37.1-176.0 km), remaining on the shallow continental shelf (<200m deep), although they moved over hundreds of kilometres (546.2 ± 154.1 km, range: 375.7-846.6 km) in waters of French Guiana and of neighbouring Surinam (**Figure 1** and **Table 1**). Four turtles remained within 50 km of Awala-Yalimapo beach, where they moved erratically in shallow waters (approximately 20 m deep). Four other turtles remained within 100 km from the beach, reaching waters approximately 50 m deep, while the last three individuals reached the edge of the continental shelf where depth is about 100 m. Among the 7 turtles that moved over significant distances, two headed West into Surinam waters where they tending to move anticlockwise before swimming along the coast towards the nesting beach. The five other turtles remained in French Guiana waters East from Awala-Yalimapo beach, tending to move clockwise until they crossed their initial path off the Maroni River. They then reached Surinam waters where they moved anticlockwise before swimming along the coast towards the nesting beach.

There was no significant relationship between turtle biometry (standard curvilinear carapace length and curvilinear carapace width) and trip duration (Spearman rank correlation, $r_S = -0.19$, $p = 0.65$; $r_S = 0.12$, $p = 0.77$, for carapace length and width, respectively, $n = 8$ turtles; 3 of the 11 turtles were not measured), dispersal range ($r_S = -0.40$, $p = 0.32$; $r_S = 0.17$, $p = 0.69$, for length and width, respectively) and total distance travelled ($r_S = -0.12$, $p = 0.78$; $r_S = 0.37$, $p = 0.37$, for length and width, respectively).

**Vertical movements**

*General characteristics*

A total of 20607 dives > 2 m was recorded from the 10 turtles, the longest dive being 28.2 min for a maximum depth of 63.9 m while the deepest dive was 83.8 m for a duration of 17.9 min (**Table 2**). Turtles performed mostly shallow (9.4 ± 9.2 m) and short (4.4 ± 3.4 min) dives with substantial variation among individuals (**Table 2**). Dives shallower than 8 m and 25 m represented 50% and 90% of the 6000 randomly sampled dives, respectively (**Figure 2a**). Dives shorter than 4 min and 10 min represented 50% and 90% of the 6000 randomly sampled dives, respectively (**Figure 2b**). Dives deeper than 40 m (n = 92) were performed by one single individual (#200101) and lasted on average 15.3 ± 3.1 min and did not significantly increase in duration with increasing depth (from 40m to 85m, ANOVA, $F_{8,91}$ = 2.02, p = 0.06). Within a dive, the time spent at the bottom lasted a mean of 1.6 ± 1.8 min, corresponding to 32.8 ± 16.0% (ranging from 1 to 90%) of the total dive duration (**Table 2**). Wiggles occurred at the bottom of most of the dives, as 50% and 90% of the dives showed less than 12, and less than 40 wiggles, respectively, with a mean of 16.3 ± 17.5 wiggles per dive (**Figure 3**). Mean surface interval was 1.4 ± 1.9 min (**Table 2**).

Due to dive depth distribution, the relationships between dive depth and other dive parameters were investigated considering dive depth ranging from 2 m to 40m with 5-m increments (individual relations were calculated with 600 random dives per turtle except #200101, n= 508 dives because of 92 dives > 40m deep; $n_{tot}$ = 5908 dives, **Figure 4**). Mean dive duration increased significantly with increasing depth class when considering either each turtle individually (Spearman rank correlation, p < 0.05 in all cases) or when considering all turtles together (Spearman rank correlation between the grand mean and depth classes, $r_S$ = 0.97, n = 10, p < 0.001). There was no significant relationship between mean bottom time and depth class when considering either each turtle individually (Spearman rank correlation, p > 0.05 in all cases) or all turtles ($r_S$ = 0.24, n = 10, p = 0.57). When considering all individuals together, both rates of descent and ascent were significantly related to dive depth ($r_S = 0.95$, n = 10, $p < 0.001$; and $r_S = 0.90$, n = 10, $p < 0.01$, respectively, Figure 4). However these relationships did not occur for several turtles, when considered individually (descent rate: #200103, #200202, #200203, #200301, #200303 and #200304; ascent rate: #200101, #200103, #200203, #200301, #200302, #20303 and #200304). Turtles descended and ascended at about 0.1 m.s$^{-1}$ for dives < 15m and regularly increased their rate of vertical travel for deeper depths with a mean maximum rate of $0.26 \pm 0.06$ m s$^{-1}$ (n=5908 dives; Figure 4).

There were significant relationships between preceding surface interval and dive duration when considering each turtle individually (Pearson rank correlation, $p < 0.05$ in all cases, n = 600 dives per turtle) except for 4 individuals (#200102, #200103, #200201, #200302, Pearson rank correlation, $p > 0.2$ in all cases, n = 600 dives per turtle). However, this relationship did not occur if all turtles are considered together ($r^2 = 0.07$, n= 6000 dives, $p = 0.23$). Similarly, there were significant relationships between post-dive surface interval and dive duration if each turtle was considered individually (Pearson rank correlation, $p < 0.05$ in all cases, n = 600 dives per turtle) except for 3 individuals (#200201, #200301, #200302, Pearson rank correlation, $p > 0.1$ in all cases, n = 600 dives per turtle). However, this relationship did not occur when considering all turtles ($r^2 = 0.09$, n= 6000 dives, $p = 0.21$).

*Dive types*

The PCA performed on dive parameters for the 20607 dives > 2 m identified 2 factors explaining 34% and 20% of the observed variance, respectively. Factor 1 and factor 2 were predominantly associated with dive duration and bottom time, respectively and enabled us to distinguish 4 main dive types (**Figures 3** and **5** and **Table 3**). *Type 1* dives were the shallowest and the shortest dives, whereas *Type 3* dives were the deepest and the longest. *Type 2* dives and *Type 4* dives had intermediate mean maximum dive depths and mean dive durations but differed in their bottom time, with *Type 2* having the shortest and *Type 4* having the longest bottom time (**Figures 3** and **5** and **Table 3**). In addition, *Type 2* dives showed the slowest ascent rate of all dives (**Table 3**).

*Diurnal pattern*

There was no significant difference between the number of dives performed during daytime and night time over the entire inter-nesting interval ($n_{day}$ = 10296 dives *versus* $n_{night}$ = 10311 dives, $\chi^2_1$ = 0.005, p > 0.05). Dives performed during daytime were significantly deeper and correspondingly longer (9.9 ± 9.2 m; 4.6 ± 3.3 min) than those performed at night (8.9 ± 9.1 m, T-test, t = 7.41, p < 0.01; 4.2 ± 3.4 min, T-test, t = 7.80, p < 0.01; **Figure 6**). Turtles performed on average 10.7 ± 3.6 dives per hour (hourly dive frequency) corresponding to a mean time spent diving of 45.0 ± 4.1 min $h^{-1}$ (**Table 2**). There was no significant difference between day and night either in terms of hourly dive frequency (Mann-Whitney, Z = 107.0, n = 10, p = 0.9) or in terms of time spent diving per hour (Mann-Whitney, Z = 123.0, n = 10, p = 0.19).

The four dive types were not equally distributed between day and night ($\chi^2_3$ = 374.6, n = 20607, p < 0.001; **Figure 7**). Shallow *Type 1*-dives predominantly occurred during night time ($\chi^2_{23}$ = 134.1, n = 11733, p < 0.001 followed by a contrast test; **Figure 7**) whereas *Type 4*-dives occurred mostly during the daytime ($\chi^2_{23}$ = 338.7, n = 3186, p < 0.001 followed by a contrast test; **Figure 7**). *Type 2* and *Type 3* were equally distributed throughout the 24-hour cycle ($\chi^2_{23}$ = 13.2, n = 5105, p > 0.05, $\chi^2_{23}$ = 8.9, n = 583, p > 0.05, respectively; **Figure 7**).

**Integrating vertical and horizontal movements**

The 7 satellite-tracked turtles for which dive records were complete throughout the inter-nesting interval spent a mean of 9.8 ± 1.3 days at sea between two consecutive nesting events (**Table 2**). This is similar to the 10.3 ± 1.4 days spent at sea by the 5 individuals only fitted with PTT (Mann-Whitney, Z = 40, n = 12, p = 0.42; **Table 1**).

All 7 turtles dived continuously throughout their inter-nesting interval but with substantial changes in dive parameters through time (**Figure 8a** and **Appendix**). Among these 7 turtles with complete dive records, 4 individuals were also successfully tracked by Argos during the entire inter-nesting interval (**Table 2**). For these 4 turtles, maximum depth and mean depth increased with distance to the beach (Spearman rank correlation, p < 0.05 in the 4 cases; **Figure 8a** and **8b**).

Similarly both indices of hourly diving effort varied significantly throughout the 7 inter-nesting intervals (Kruskal-Wallis, p < 0.05 in all cases; **Figure 8c and Appendix**) except for #200202, for which time spent diving per hour did not vary significantly ($H_{16, 191}$ = 19.5, n = 192, p = 0.24). For each inter-nesting interval, two phases were identified according to the way the time spent diving per hour varied through time (**Figure 8c and Appendix**). Indeed, the time spent diving per hour was significantly higher during phase 1 (47.3 ± 3.0 min $h^{-1}$) than during phase 2 (31.3 ± 6.7 min $h^{-1}$, Mann-Whitney, Z = 77.0, n = 7 turtles, p < 0.01; **Figure 8c and Appendix**). Yet the hourly dive frequency did not differ significantly between phase 1 (9.7 ± 3.0 dives $h^{-1}$) and phase 2 (12.4 ± 2.6 dives $h^{-1}$, Mann-Whitney, Z = 39.0, n = 7 turtles, p = 0.09; **Figure 8c and Appendix**).

Phase 1 lasted a mean of 7.2 ± 1.3 days (i.e about 75% of the inter-nesting interval), with significant differences among individuals in the time spent diving per hour (47.2 ± 2.9 min $h^{-1}$, ANOVA, $F_{6,104}$ = 14.0, n = 7 turtles, p < 0.001) due to a particularly low value for one individual (#200201, 40.9 ± 4.4 min $h^{-1}$, post-hoc Tuckey test) compared to the 6 others which did not differ (48.3 ± 1.4 min $h^{-1}$, ANOVA, $F_{5,86}$ = 1.6, n = 6 turtles, p = 0.17).

Similarly, during that first phase, the hourly dive frequency showed substantial differences among individuals (9.7 ± 3.0 dives h$^{-1}$, ANOVA, $F_{6, 104}$ = 12.4, n = 7 turtles, p < 0.001). Turtles performed deeper, but fewer, dives as they moved away from the coast **(Figure 8b, 8c and Appendix)**. Turtles which remained within 50 km of the coast (#200103, #200301) performed 13.0 ± 1.9 dives h$^{-1}$, mostly short and shallow *Type 1*-dives (47.0 ± 18.8% of their recorded dives). Turtle #200102, which dispersed between 50-100 km from the beach, performed 8.8 ± 2.4 dives h$^{-1}$, predominantly dives of intermediate depth of *Type 2* and *Type 4* (40.5 ± 20.3% and 31.9 ± 20.8% of her recorded dives, respectively). Finally, #200101 moved over more than 100 km from the beach and performed the fewest dives (5.0 ± 3.0 dives h$^{-1}$) but most of them (69.1 ± 55.5%) were long and deep *Type 3*-dives.

Phase 2 lasted a mean of 2.6 ± 1.8 days (i.e about 25% of the inter-nesting interval). For each turtle, this phase was highly variable in terms of hourly diving effort (**Figure 8c** and **Appendix**). However, the time spent diving per hour and the hourly dive frequency did not differ significantly among turtles and averaged 31.3 ± 6.7 min h$^{-1}$ and 12.5 ± 2.6 dives h$^{-1}$ respectively (Kruskal-Wallis, $H_{6,37}$ = 6.7, p = 0.35, and $H_{6,37}$ = 9.18, p = 0.16, respectively, n = 7 turtles). Phase 2 was predominantly associated with short and shallow *Type 1*-dives suggesting that the turtles were generally swimming at the surface when moving back to the nesting beach (**Figure 8b**) but also with a non-negligible proportion (13.1 ± 3.9%) of *Type 4*-dives.

## DISCUSSION

The diving behaviour of leatherback turtles during the nesting season has been widely studied (Eckert et al. 1986, 1989, 1996, 2002, in press; Keinath & Musik 1993, Southwood et al. 1999, Hays et al. 2004a, Reina et al. 2005, Wallace et al. 2005, Myers & Hays in press), but to date, only one study proposed the concurrent analysis of diving behaviour with dispersal patterns assessed by satellite telemetry in this species during the nesting season (Eckert et al. in press). Our study shows that leatherback turtles nesting in French Guiana have a wide range of dispersal over the continental shelf, moving over hundreds of kilometres in waters of French Guiana and neighbouring Surinam and show a bathymetry-constrained dive pattern.

## General dispersal patterns

In French Guiana, leatherbacks explore three main zones of the continental shelf, ranging from the costal zone (within 50 km from the coast), the neritic zone (within 100 km from the coast) to the edge of the continental shelf. There are important individual differences in the distance travelled and in the diving pattern during the inter-nesting interval, yet, leatherbacks spend a similar time at sea regardless of their dispersal effort. Note here that our individuals show inter-nesting intervals of similar duration and dispersal range to those of leatherbacks in other Atlantic nesting sites, whether equipped or not with externally-attached instruments (Miller 1997, Georges et al. in press). This indicates that the instruments have a negligible effect on that aspect of the turtles' behaviour (see Fossette et al. submitted). The observed individual variations are apparently not related to the individual size. Since sea turtles grow continuously throughout their life-span (Chaloupka & Musick 1997) this suggests that the duration of the inter-nesting interval and the dispersal pattern are not age-related. It might rather be related to individual body condition or to oceanographic conditions (see Gaspar et al. 2006) but identifying the actual causes of the inter-individual variability was beyond the scope of this study.

## General diving patterns

In French Guiana, leatherback diving behaviour appears to be restricted by the bathymetry of the continental shelf, with 90% of the dives being shallower than 25 m and shorter than 10 min. Dive duration, thus, never exceeds the Aerobic Dive Limit (ADL) estimated between 33 and 67 min (Southwood et al. 1999, Hays et al. 2004a, Wallace et al. 2005). This aerobic diving behaviour is supported by the weak or absent relationship between post-dive surface interval and dive duration. Similarly, leatherbacks do not appear to anticipate the duration of their dive as suggested by the weak or absent relationship between preceding surface interval and dive duration. Such a short and shallow diving activity is similar to that reported for turtles of the Eastern Pacific (Southwood et al. 1999, Reina et al. 2005, Wallace et al. 2005) and the China Sea (Eckert et al. 1996) but differs from the deep pelagic dives performed by leatherbacks in the Caribbean Sea (Eckert et al. 1989, Eckert 2002, Hays et al. 2004a). Descent and ascent rates are comparable, yet lower, in French Guiana (0.13 ± 0.09 m $s^{-1}$ and 0.11 ± 0.11 m $s^{-1}$, respectively) than in the Eastern Pacific (descent rate: 0.15 ± 0.06 m $s^{-1}$, ascent rate: 0.20 ± 0.11 m $s^{-1}$ estimated from Reina et al. 2005) and China Sea (descent rate: 0.20 ± 0.05 m $s^{-1}$, ascent rate: 0.20 ± 0.04 m $s^{-1}$ estimated from Eckert et al. 1996) but much lower than those estimated in the Caribbean Sea (descent rate: 0.41 ± 0.18 m $s^{-1}$, ascent rate: 0.31 ± 0.01 m $s^{-1}$, estimated from Eckert 2002). As in Costa Rica (Reina et al. 2005), French Guiana leatherbacks seem to stroke continuously when moving vertically as suggested by the similar vertical rates during descent and ascent. However, in Costa Rica leatherback turtles maintain very low energy expenditure during the inter-nesting interval (Wallace et al. 2005) by laying motionless on the seafloor (Reina et al. 2005). In contrast, leatherbacks from French Guiana spend almost one third of their time at the bottom of the dive, where they perform numerous, substantial wiggles, suggesting that they actively swim throughout the dive. This is supported by direct measurements of the actual swim speed in leatherbacks from French Guiana (H.T. Pers. Comm.). Additionally, when swimming back to the nesting beach, leatherbacks from French Guiana move to the proximity of the shore in the last days of the inter-nesting interval, presumably in anticipation of egg laying. Such patterns may also allow gravid turtles to cope with potential early egg-laying. The observation of a similar behaviour in leatherback turtles nesting in Gabon (Georges et al. in press), on the other side of the Atlantic basin, is consistent with the fact that the duration of the inter-nesting interval is restricted by the timing of egg-maturation. In short, leatherbacks appear to adopt at least two strategies during their inter-nesting intervals according to their nesting site: the "Pacific strategy" consists in reducing swim activity to limit energy expenditure between two consecutive ovi-positions, and the "Atlantic strategy" where leatherbacks dive and swim almost continuously throughout the inter-nesting interval while dispersing extensively, probably for feeding (Eckert et al. 1989, in press, Myers & Hays in press, this study). These two strategies may be linked with the local oceanographic conditions, with local food availability probably shaping the behaviour in the different nesting sites, as also reported in green turtles (Hays et al. 2002).

**Combined analysis**

The bathymetry of the Guiana's continental shelf is not precisely known because of the continual influence of Amazon-derived mud banks on its morphology (Anthony & Dolique 2004) but in a general sense, depth increases gradually with distance from the coast (**Figure 1**). In our study, leatherback turtles dived deeper as they moved away from the coast, implying that they tend to follow the topography of the seabed as has been suggested for their relatives from the China Sea (Eckert et al. 1996). This is supported by occasional, direct observations of well-adhered, but fresh, mud on the carapace of nesting leatherbacks in French Guiana (personal observations), suggesting that the mud was acquired some appreciable time before hauling out.

One striking finding of our study is that, despite the high inter-individual variability in their dispersal effort, all leatherback individuals spend the same amount of time diving (80% of their time spent at sea). Indeed, we found that during the first 75% of the inter-nesting interval leatherbacks dived on average 47 min per hour with a striking consistency among individuals, whereas they showed a lower and more variable diving effort thereafter when returning to the coast. Similarly Southwood et al. (1999) identified two phases in the diving behaviour of leatherbacks in Costa Rica, with a first phase showing relatively deep and long dives compared to the rest of the inter-nesting interval. Additionally, we found that the shapes of the dives changed according to the two phases defined above and from the distance to the nesting beach.

*Close to the nesting beach*

Immediately after leaving the beach and when swimming back to their nesting site, leatherbacks mainly perform shallow (5m-deep) *Type 1*-dives. These dives are the most abundant in all inter-nesting intervals. Comparable dive types have been reported for green (*Chelonia mydas*) and loggerhead (*Caretta caretta*) turtles (Minamikawa et al. 1997, Hochscheid et al. 1999, Hays et al. 2001, Houghton et al. 2002) as well as for leatherbacks (Eckert et al. 1996, Southwood et al. 1999, Eckert 2002, Reina et al. 2005, Wallace et al. 2005), and have been interpreted as travelling dives (Papi et al. 1997, Eckert 2002, Reina et al. 2005). *Type 1*-dives may contribute to optimal travelling behaviour, as animals avoid unstable conditions at the sea surface (e.g pitching and rolling due to wave action) as well as reducing surface drag so that transport costs are minimized (Minamikawa et al. 2000). *Type*

*1*-dives mostly appeared at night with peaks at the beginning of the night (07:00 pm - 09:00 pm) and later, around 02:00 am suggesting that leatherbacks may travel more at these times. By contrast, Eckert (2002) suggested that in the Caribbean Sea, leatherbacks may perform most of their horizontal movements during the middle of the day.

*In the coastal zone*

Within the first 12 hours of departure, the shape of dives changes according to the dispersal mode: in addition to the travel-like *Type 1*-dives, leatherbacks remaining in the costal zone (within 50 km from the coast) also perform dives of similar shape but slightly deeper (15m-deep, *Type 2*-dives), suggesting that travelling also occurs at deeper depths. These *Type 2*-dives have a particularly slow (0.08 m s$^{-1}$) and progressive ascent to the surface compared to other dives, similar to the 'S-shaped' dives described for green and loggerhead turtles by Minamikawa et al. (1997), Hochscheid et al. (1999), Hays et al. (2001) and Houghton et al. (2002). In green and loggerhead turtles, the slow and progressive ascent has been interpreted as a passive locomotion due to positive buoyancy. Contrary to green and loggerhead turtles, however, the small lung volume of leatherbacks probably does not function for buoyancy control (Minamikawa et al. 1997). This progressive ascent may also correspond to prey searching and/or capture as suggested in other diving predators (e.g. Ropert-Coudert et al. 2001). In the case of leatherback turtles in French Guiana, the nesting season coincides with the peak of Amazonian influence, resulting in rings of Amazonian waters crossing the continental shelf (Froidefond et al. 2002), enhancing biological productivity. This includes the production of gelatinous plankton whose relatively large individual size and quantity may have a potentially significant contribution to the marine food web on the continental shelf (Fromard et al. 2004), particularly for leatherbacks which predominantly feed on jellyfish (James & Herman 2001). Massive stranding (up to several hundred individuals on the 2km long beach) of jellyfish of genus *Rhizostoma* sp. and occasionally *Aurelia* sp. are regularly observed on Awala-Yalimapo beach during the nesting season (personal observation). Both species have been reported as common prey for leatherbacks in the Atlantic (James & Herman 2001) supporting the hypothesis that in French Guiana leatherbacks may encounter profitable food conditions during their inter-nesting intervals, as reported for neighbouring Caribbean sites (Eckert et al. 1989, press, Myers & Hays in press).

*In the neritic zone and the edge of the continental shelf*

When moving further on the continental shelf, leatherbacks maintained *Type 2*-dives but also performed 10m-deep *Type 4*-dives until they reached the edge of the continental shelf, where *Type 4*-dives were replaced by *Type 3*-dives. *Type 3* and *Type 4* dives are 'W-shaped' dives (Wilson 1995) characterised by a relatively long bottom phase during which numerous wiggles of several meters amplitude (> 2m) occur. These wiggles are commonly interpreted as corresponding to prospecting and foraging behaviours (e.g. Schreer et al. 2001), supporting the hypothesis that leatherbacks may attempt to feed throughout the continental shelf. *Type 4*-dives mostly occurred during the day, suggesting that if they were associated with prospecting and foraging, leatherbacks may feed on gelatinous prey during daytime. If that was the case, the Guyanese situation would differ from the Caribbean, where leatherbacks are supposed to be nocturnal feeders (Eckert et al. 1989). Such a difference may explain why leatherbacks travel predominantly during the night in French Guiana (this study) but during the day in the Caribbean Sea (Eckert 2002). Inter-site differences may also result from local bathymetry. In the deep Caribbean Sea, jellyfish may only be accessible to leatherbacks at night, as they migrate to the surface, whereas in French Guiana, the relatively shallow water depths of the continental shelf restrict jellyfish vertical movement both during day- and night-time. Further investigations are required to clarify the behaviour and abundance of jellyfish in Guyanese waters, since nocturnal shallow *Type 1*-dives may also contribute to nocturnal foraging when jellyfish migrate to the surface. In addition, if gravid leatherbacks of French Guiana feed during the inter-nesting intervals, this could explain why they are on average heavier than their relatives from other nesting sites (Georges & Fossette 2006). Wallace et al. (2005) hypothesized that gravid leatherbacks in Costa Rica do not actively forage during the nesting season as they rarely approach their Aerobic Diving Limits (ADLs). Leatherbacks in French Guiana also dive within their ADL yet our results suggest that they may feed opportunistically rather than optimally (Thompson & Fedak 2001).

Thus our combined analysis of dispersal and diving data suggests that leatherbacks may compensate for high reproductive expenditure by extensively prospecting for food during the inter-nesting intervals. Direct evidence is however, required to confirm this. It could be tested using underwater video, or by measuring the energy balance or body mass changes during the inter-nesting intervals. Further investigations are also required to clarify the observed inter-individual variability we found in the dispersal and diving patterns, and their implications in terms of individual fitness.

**Conservation implications**

If leatherback turtles forage or actively prospect for food during the nesting season, they may be exposed to high risks of interactions with fisheries, particularly in the Guiana shield where international fisheries operate (Charuau 2002). For some sea turtles populations which remain close to their nesting beach during the breeding season exploiting restricted foraging grounds, focused conservation efforts on such areas can prove tremendously successful (e.g. Hays 2004b). However in French Guiana, leatherback turtles move far away from their nesting beach during a single inter-nesting interval and cross international borders to move into Surinam waters. This emphasises the need for regional conservation strategies and international control of fishing practices in this area which is heavily exploited by two local fisheries (Charuau 2002, Georges et al. in press). In addition, leatherbacks exploit the entire water column, following the bathymetry of the continental shelf. As a consequence, trawlers operating in French Guiana and Surinam waters should adopt appropriate fishing gear (e.g. Lutcavage et al. 1997, Epperly 2003) to reduce accidental captures of endangered leatherback turtles.

*Acknowledgements.* We are grateful to the Ministry of Ecology and Sustainable Development and the Direction Régionale de l'Environnement-Guyane in French Guiana. We thank all participants of sea turtle monitoring programmes developed in Awala-Yalimapo beach (Réserve Naturelle de l'Amana, Kulalasi and WWF) for logistical help in the field. We also thank D. Grémillet for his comments on the first draft. S. Fossette was supported by a studentship from the French Ministry of Research. Funding was provided by grants to Y. Le Maho from European FEDER program, to H. Tanaka from Research Fellowships of the Japan Society for the Promotion of Science for Young Scientists, from the Japan Society for the Promotion of Science (14405027 and 15255003) and from the MEXT Grant-in-Aid for the 21st Century COE Program "Neo-Science of Natural History" Program at Hokkaido University. This study was carried out under CNRS institutional license (B67-482 18).

Table 1. Summary of the inter-nesting movements performed by 11 Argos tracked leatherback turtles nesting in French Guiana in 2001, 2002 and 2003. Six turtles were simultaneously equipped with TDR (+ ; see Table 2).

Table 2. Summary of diving behaviour (for dives > 2m in depth) in 10 leatherback turtles nesting in French Guiana in 2001, 2002 and 2003. Values are expressed as mean ± SD (max value). Six turtles were simultaneously tracked with PTTs during one complete inter-nesting interval (+; see Table 1). * Mean trip duration was calculated only from the seven complete diving records.

Table 3. Statistics (mean ± SD) of parameters for the 4 types of dive in 10 gravid leatherback turtles during their interesting interval in French Guiana during the nesting seasons 2001, 2002, and 2003. For each parameters and each dive type, the mean values were always significantly different to each other (post-hoc Tukey test, P< 0.05 in all cases). A PCA identified two axes associated with dive duration (axe 1) and bottom time (axe 2) shown in bold in the table.

Fig.1. Inter-nesting movements performed by 11 gravid leatherback turtles nesting in French Guiana in relation to bathymetry during the nesting seasons 2001 (a), 2002 (b) and 2003 (c).

Fig. 2. Frequency distribution of (a) dive depth and (b) dive duration (n = 6000 dives) in 10 gravid leatherback turtles during their inter-nesting interval in French Guiana in 2001, 2002 and 2003.

Fig. 3. (a) Dive profile of a gravid leatherback turtle (#200102) during a simple inter-nesting interval in French Guiana (see Table 1). (b), (c), (d) Enlarged profiles during 6 hours at the start, middle and end of the inter-nesting interval respectively. (e), (f), (g) Enlarged profile during 2 hours illustrating '*Type 4*', '*Type 2*' and '*Type 1*' dives respectively (see results for details). (1) shows a classical wiggle pattern

Fig. 4. Relationships between dive depth and (a) dive duration, (b) proportion of time spent at the bottom of dives and (c) rates of descent and ascent in 10 gravid leatherback turtles during their inter-nesting interval in French Guiana in 2001, 2002 and 2003. Individual relations were calculated with 600 random dives per turtle (except #200101, n=508 dives, see results) from which the mean relationship was calculated (mean ± SD, open circles, n = 10 individuals).

Fig. 5. Schematic representation of the four dive types in 10 gravid leatherback turtles during their inter-nesting interval in French Guiana in 2001, 2002 and 2003 (see Table 3 for details). Dives chronologically consisted in a descent phase, a bottom time, an ascent phase and a post-dive surface interval (shown here for *Type 3* as an example).

Fig. 6. Distribution of dives in relation to time of the day and dive depth in 10 gravid leatherback turtles during their inter-nesting interval in French Guiana in 2001, 2002 and 2003 (n = 6000 dives). The solid lines along the Z axis show night time.

Fig. 7. Hourly distribution of each dive type (indicated by the number on the right hand side; see fig.5) performed by 10 gravid leatherback turtles during their inter-nesting interval in French Guiana in 2001, 2002 and 2003 (n = 20607 dives). The solid lines along the X axis show night time.

Fig. 8: Diving behaviour and diving effort throughout the inter-nesting interval performed by 3 gravid leatherback turtles concurrently monitored with PTT and TDR during the 2001 nesting season in French Guiana. For clarity, figure presents data for only 3 leatherback turtles considered as representative of the 3 dispersion patterns (coastal, neritic, edge of the continental shelf) observed in this study. Additional individual data are presented in Appendix. (a) Dive profile and mean depth (solid grey line), (b) daily frequency of each dive type and distance from the beach (black line), (c) diving effort, the 2 paralleled lines indicate the transition between phase 1 and phase 2 (see results).

Types of dive: 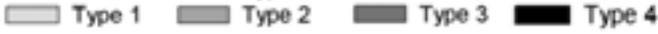

Appendix: Diving behaviour and diving effort throughout the inter-nesting interval performed by 4 gravid leatherback turtles concurrently monitored with PTT and TDR during the 2002 and 2003 nesting seasons in French Guiana. (a) Dive profile and mean depth (solid grey line), (b) daily frequency of each dive type and distance from the beach (black line), (c) diving effort, the 2 paralleled lines indicate the transition between phase 1 and phase 2 (see results). Distance to the beach is not shown for 3 turtles which were not equipped with PTTs. Types of dive: 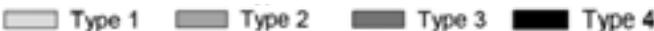

| Turtles ID no. | TDR | Departure time | Trip duration (d) | Dispersion range (km) | Total travelled distance (km) | Outbound duration (days) / distance (km) | Middle stage duration (days) /distance (km) | Return duration (days) /distance (km) | No. of locations |
|---|---|---|---|---|---|---|---|---|---|
| 200101 | + | 16 May 2001, 03:30 | 10.8 | 154.8 | 534.9 | 5.2 / 264.6 | 0.0 / 0.0 | 5.5 / 270.3 | 56 |
| 200102 | + | 22 May 2001, 06:00 | 9.9 | 70.4 | 399.1 | 3.0 / 188.4 | 1.9 / 58.0 | 5.0 / 222.7 | 101 |
| 200103 | + | 29 May 2001, 02:45 | 10.1 | 54.1 | 464.2 | 2.1 / 134.3 | 4.7 / 269.4 | 3.3 / 60.5 | 43 |
| 200104 | - | 29 June 2001, 01:30 | 11.3 | 146.5 | 771.6 | 3.8 / 318.3 | 0.0 / 0.0 | 7.5 / 453.3 | 114 |
| 200105 | - | 22 July 2001, 07:30 | 10.0 | 75.9 | 846.6 | 3.1 / 281.6 | 2.9 / 260.3 | 4.1 / 304.8 | 87 |
| 200106 | - | 26 Apr 2001, 00:30 | 8.3 | 53.7 | 375.7 | 0.5 / 34.9 | 6.7 / 232.9 | 1.9 / 107.9 | 42 |
| 200107 | - | 01 June 2001, 03:30 | 11.8 | 78.5 | 598.4 | 2.1 / 113.1 | 2.4 / 58.5 | 7.3 / 426.8 | 56 |
| 200203 | + | 16 May 2002, 02:15 | 10.6 | 99.4 | 411.8 | 2.1 / 148.0 | 0.0 / 0.0 | 8.5 / 263.8 | 47 |
| 200204 | - | 03 June 2002, 03:20 | 10.2 | 176.0 | 641.8 | 2.8 / 193.7 | 0.0 / 0.0 | 7.4 / 448.1 | 51 |
| 200301 | + | 05 May 2003, 21:26 | 9.3 | 37.1 | 464.9 | 0.6 / 51.5 | 7.0 / 325.4 | 1.8 / 88.0 | 84 |
| 200303 | + | 07 May 2003, 02:33 | 10.2 | 47.9 | 498.6 | 2.5 / 126.1 | 6.2 / 293.6 | 1.4 / 78.9 | 134 |
| Mean ± SD | | | 10.2 ± 0.9 | 90.4 ± 47.7 | 546.2 ± 154.1 | 2.5 ± 1.3 / 162.2 ± 92.3 | 2.9 ± 2.8 / 136.2 ± 137.6 | 4.9 ± 2.6 / 247.7 ± 150.9 | 74.1 ± 31.7 |

| Turtle ID no. | ARGOS PTT | Departure time | Trip duration (d) | No. of dives | Dive depth (m) | Dive duration (min) | Bottom time / dive duration (%) | Post-dive surface interval (min) | Diving effort Dives.h$^{-1}$ | Min.h$^{-1}$ |
|---|---|---|---|---|---|---|---|---|---|---|
| 200101 | + | 16 May 2001, 00:26 | 10.8 | 1804 | 17.0 ± 19.7 (83.8) | 6.2 ± 5.9 (28.2) | 33.7 ± 15.8 (88.5) | 2.1 ± 3.6 (43.1) | 7.2 ± 6.1 | 42.1 ± 10.5 |
| 200102 | + | 22 May 2001, 02:58 | 9.9 | 2271 | 10.2 ± 7.0 (30.2) | 4.5 ± 2.4 (17.0) | 33.6 ± 14.6 (84.3) | 1.6 ± 1.9 (47.7) | 9.8 ± 2.6 | 42.8 ± 8.8 |
| 200103 | + | 28 May 2001, 23:46 | 9.1 | 3539 | 5.7 ± 3.6 (18.4) | 3.1 ± 1.9 (12.0) | 29.0 ± 13.7 (86.2) | 0.8 ± 1.3 (41.2) | 14.9 ± 4.1 | 45.3 ± 8.2 |
| 200201 | − | 30 Apr 2002, 23:30 | 12.1 | 2373 | 11.0 ± 8.8 (36.6) | 4.4 ± 2.8 (19.0) | 25.8 ± 13.6 (88.2) | 2.6 ± 2.9 (48.0) | 8.6 ± 2.3 | 37.3 ± 8.4 |
| 200202 | − | 02 May 2002, 22:55 | 8.2 | 1917 | 10.9 ± 7.8 (27.4) | 5.0 ± 3.4 (19.4) | 30.4 ± 15.2 (88.4) | 0.9 ± 0.9 (11.3) | 10.0 ± 4.5 | 48.7 ± 7.1 |
| 200203 | + | 15 May 2002, 23:34 | incomplete (4.6) | 1945 | 8.5 ± 9.4 (38.8) | 2.7 ± 3.1 (13.0) | 38.0 ± 12.7 (86.9) | 0.7 ± 1.0 (7.8) | 18.0 ± 19.6 | 46.9 ± 2.0 |
| 200301 | + | 05 May 2003, 22:43 | 9.3 | 2587 | 5.2 ± 2.7 (13.8) | 3.9 ± 2.8 (14.2) | 33.9 ± 15.6 (93.5) | 1.0 ± 1.4 (37.3) | 11.9 ± 4.0 | 45.1 ± 11.1 |
| 200302 | − | 06 May 2003, 22:05 | 9.2 | 2384 | 8.2 ± 5.9 (25.5) | 4.2 ± 2.4 (14.0) | 45.4 ± 17.8 (91.0) | 1.1 ± 1.2 (29.1) | 11.4 ± 2.9 | 41.7 ± 11.6 |
| 200303 | + | 09 May 2003, 01:02 | incomplete (4.0) | 679 | 10.2 ± 3.5 (16.6) | 7.3 ± 3.8 (19.2) | 29.6 ± 17.1 (86.8) | 1.3 ± 0.9 (5.0) | 7.0 ± 1.2 | 50.4 ± 2.2 |
| 200304 | − | 09 May 2003, 01:03 | incomplete (5.8) | 1108 | 14.8 ± 7.2 (32.3) | 6.5 ± 3.5 (16.9) | 24.6 ± 13.7 (79.0) | 1.1 ± 0.8 (7.2) | 7.7 ± 1.7 | 49.7 ± 2.4 |
| All turtles | | | 9.8 ± 1.3* | 20607 | 9.4 ± 9.2 (83.8) | 4.4 ± 3.4 (28.2) | 32.8 ± 16.0 (93.5) | 1.4 ± 1.9 (48.0) | 10.7 ± 3.6 | 45.0 ± 4.1 |

| | No. of dives | Frequency (%) | Depth (m) | **Duration (min)** | Descent rate (m/s) | Ascent rate (m/s) | **Bottom time (% of dive duration)** | No. of wiggles at bottom | Depth amplitude at bottom (m) | Post-dive surface interval (min) |
|---|---|---|---|---|---|---|---|---|---|---|
| Type 1 | 11733 | 56.9 | 4.9 ± 3.5 | **2.3 ± 1.6** | 0.11 ± 0.06 | 0.12 ± 0.06 | **30 ± 11** | 7.9 ± 5.5 | 0.6 ± 0.4 | 1.1 ± 5.5 |
| Type 2 | 5105 | 24.8 | 15.5 ± 7.1 | **6.8 ± 2.5** | 0.16 ± 0.12 | 0.08 ± 0.05 | **23 ± 9** | 14.3 ± 8.1 | 1.7 ± 0.7 | 1.7 ± 1.2 |
| Type 3 | 583 | 2.8 | 40.2 ± 18.9 | **13.1 ± 4.1** | 0.20 ± 0.07 | 0.20 ± 0.52 | **40 ± 17** | 54.1 ± 33.1 | 4.6 ± 2.3 | 3.9 ± 4.9 |
| Type 4 | 3186 | 15.5 | 10.7 ± 6.9 | **6.4 ± 3.0** | 0.16 ± 0.11 | 0.12 ± 0.09 | **59 ± 11** | 38.7 ± 17.8 | 2.0 ± 2.1 | 1.4 ± 1.0 |
| 1-way Anova p value | - | - | $F_{3,20607}$=8294 P < 0.001 | **$F_{3,20607}$=9149 P < 0.001** | $F_{3,20607}$=713 P < 0.001 | $F_{3,20607}$=348 P < 0.001 | **$F_{3,20607}$=8177 P < 0.001** | $F_{3,20607}$=9602 P < 0.001 | $F_{3,20607}$=4485 P < 0.001 | $F_{3,20607}$=94 P < 0.001 |

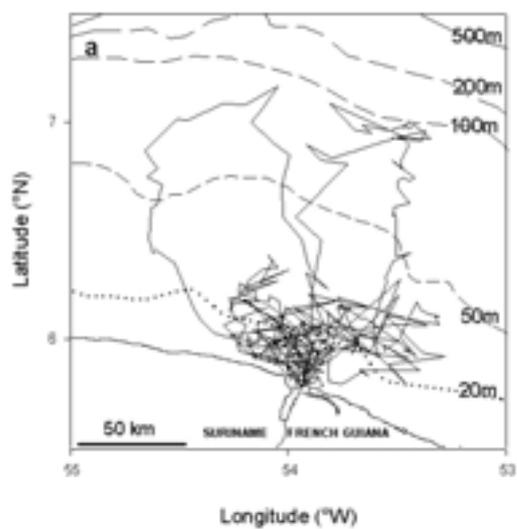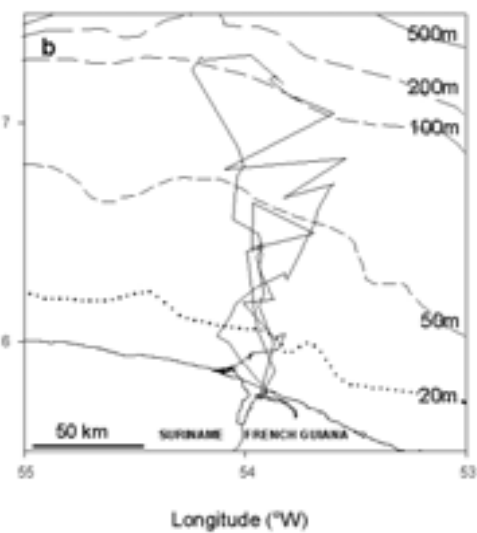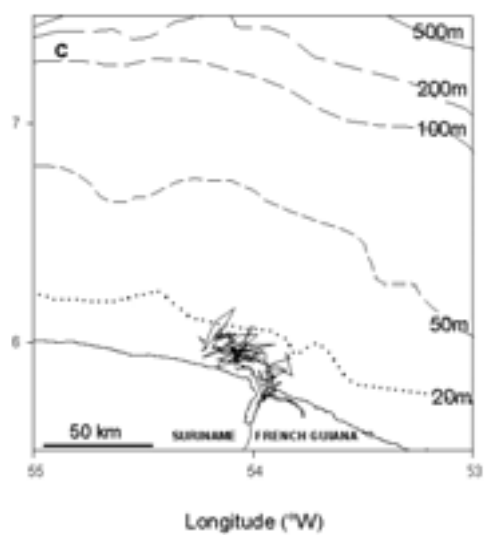

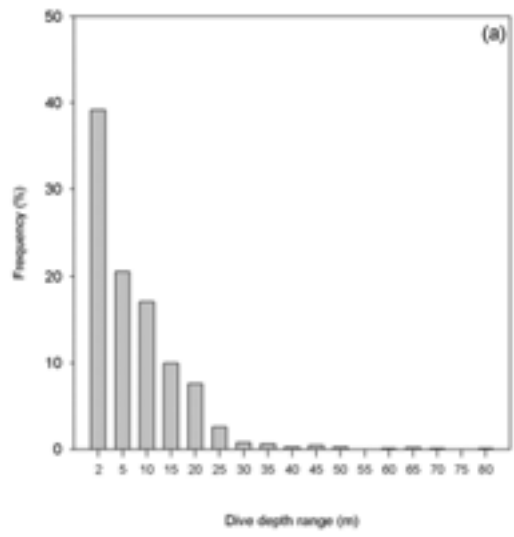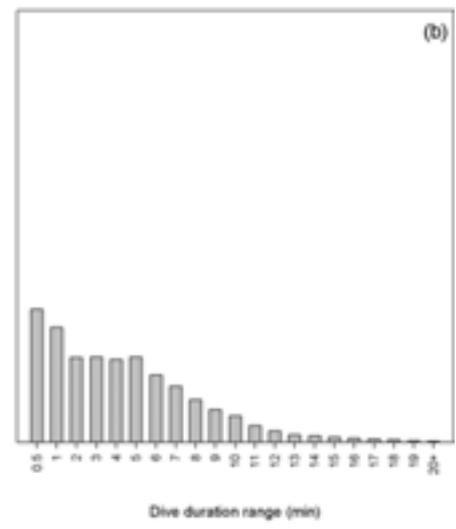

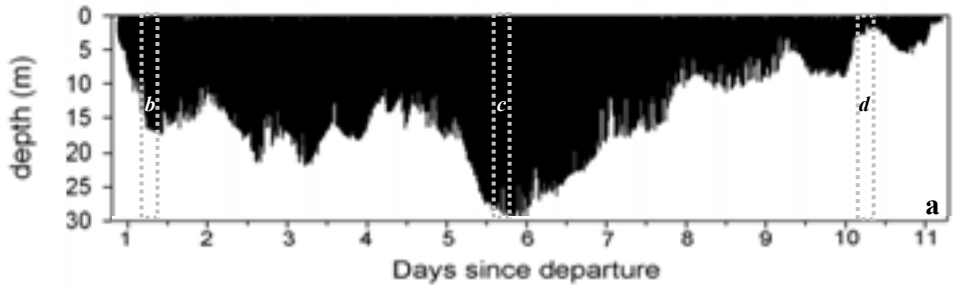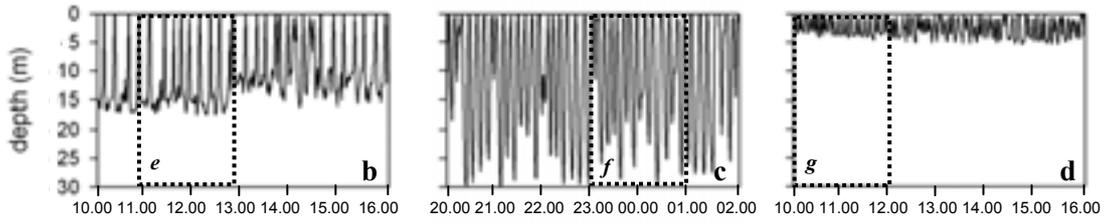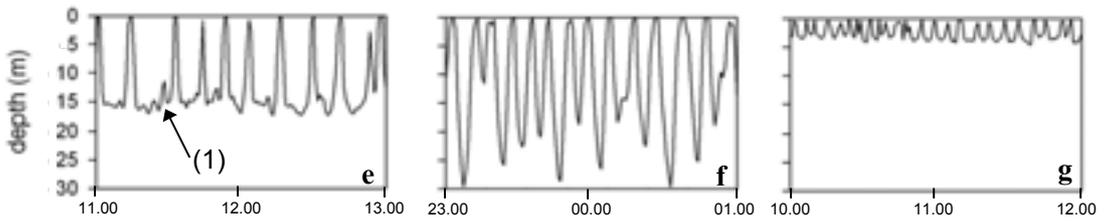

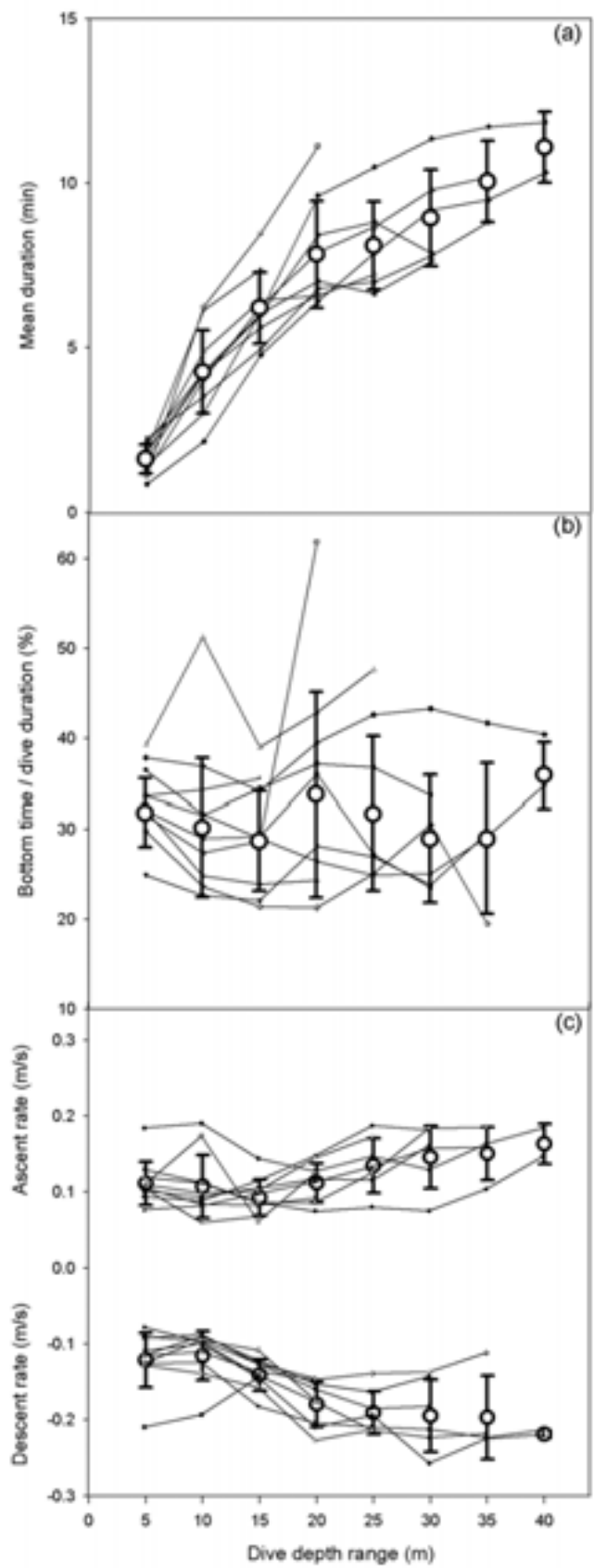

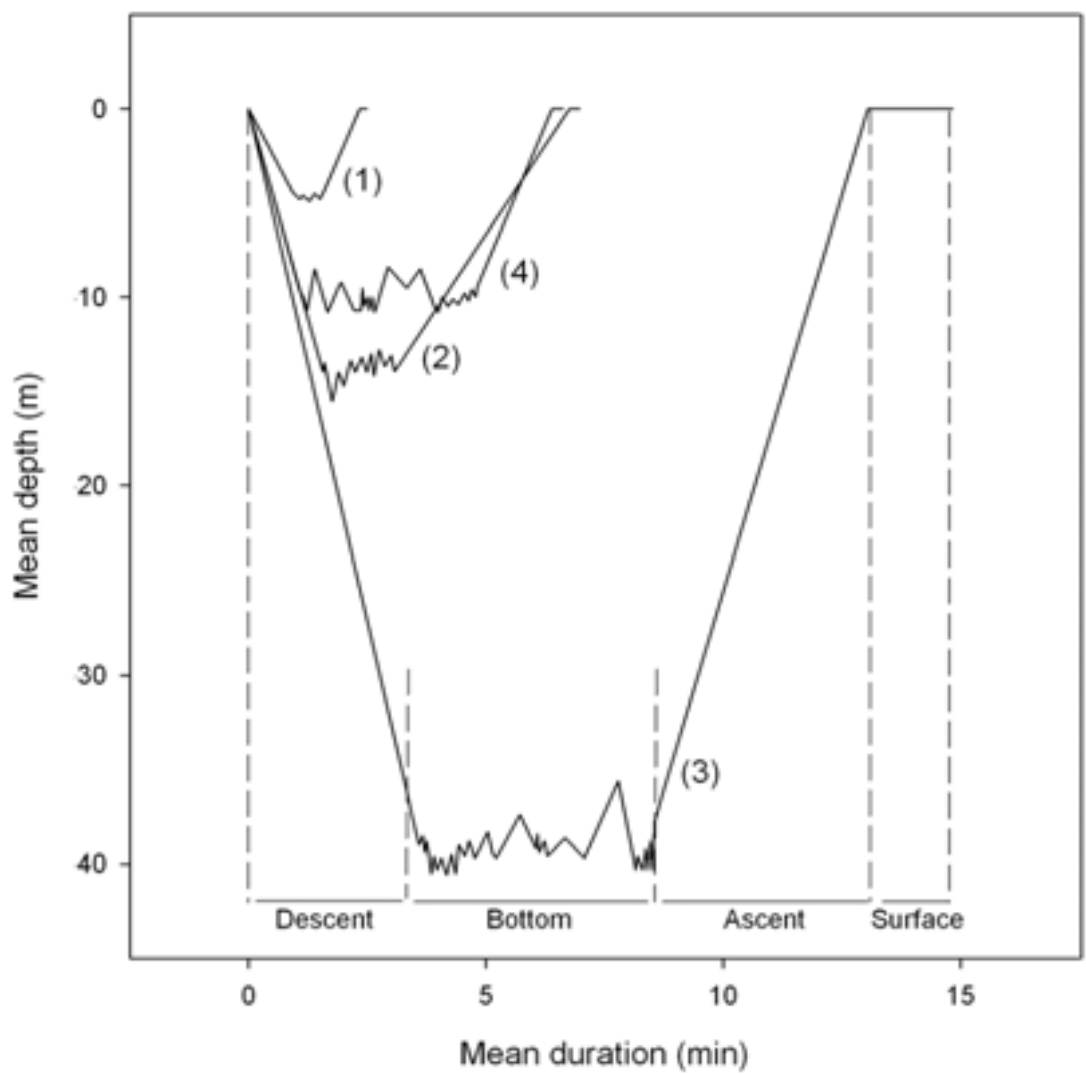

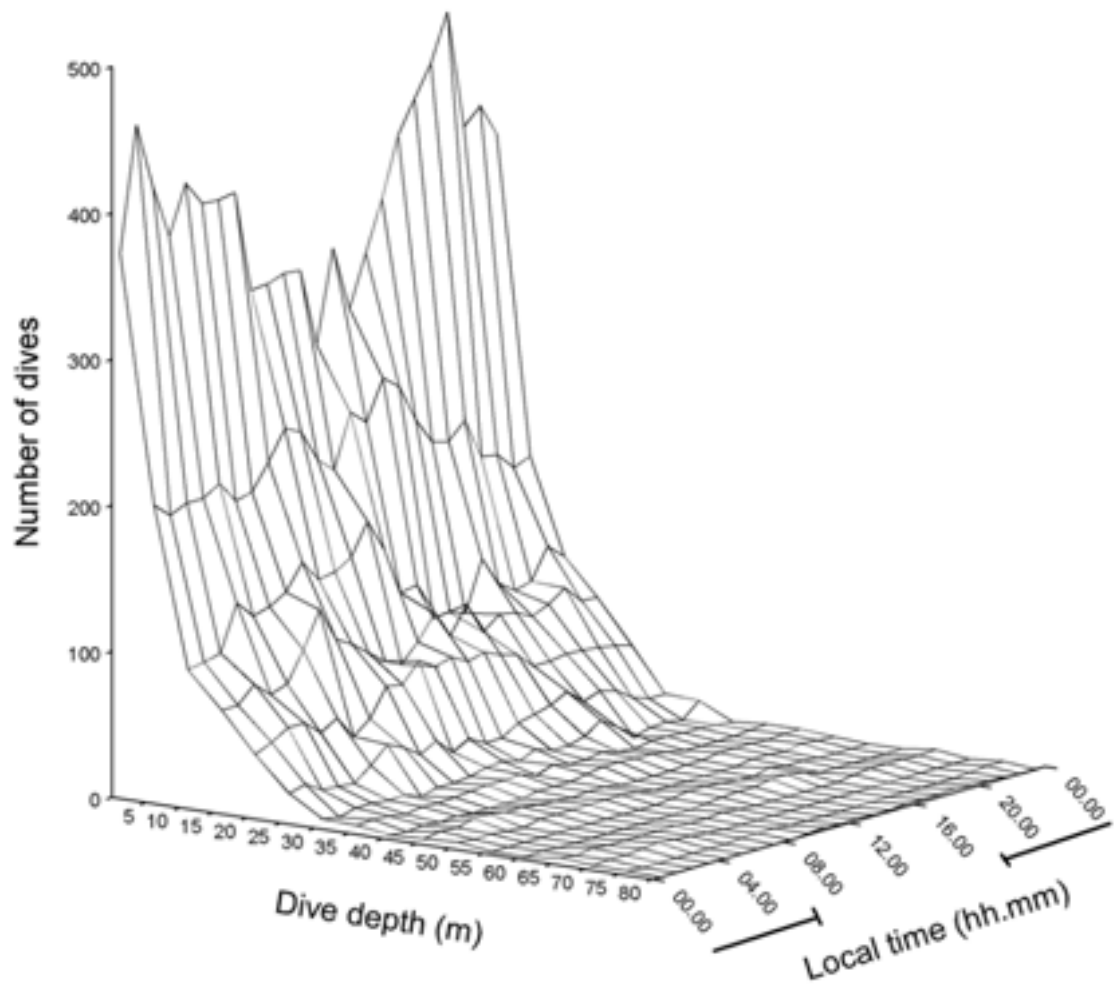

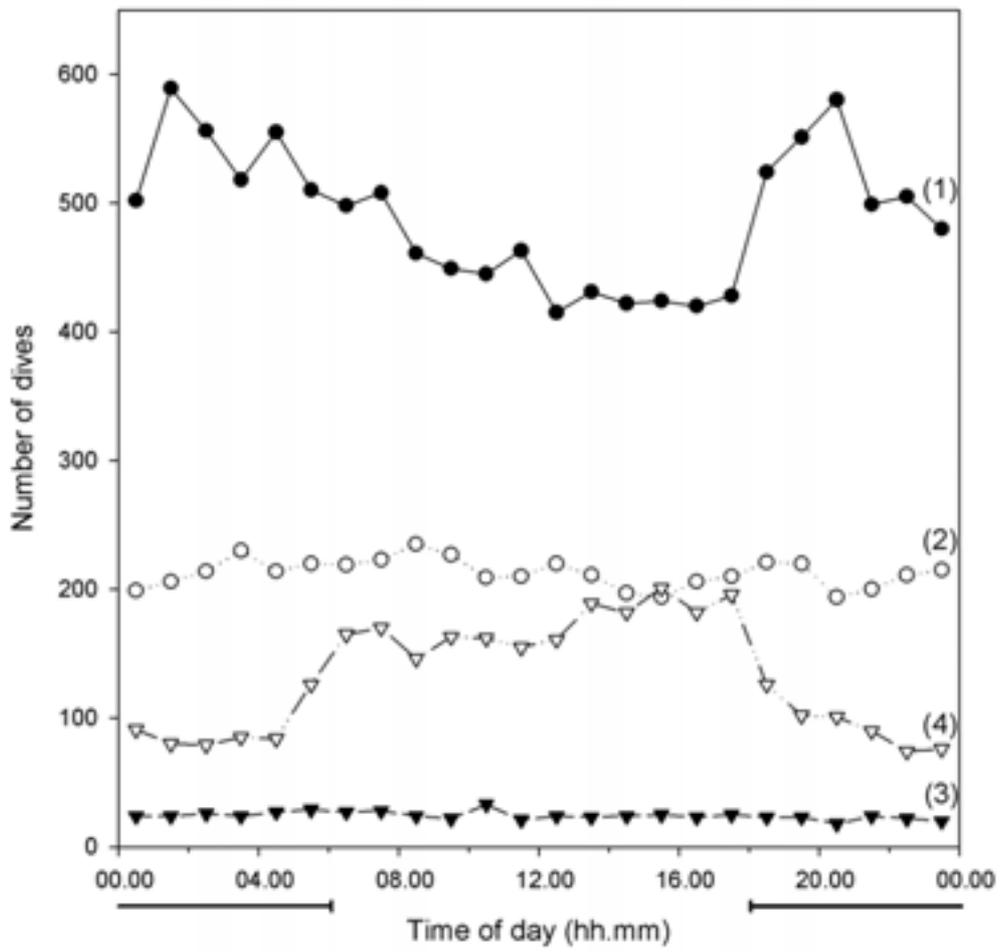

200101 200102 200103

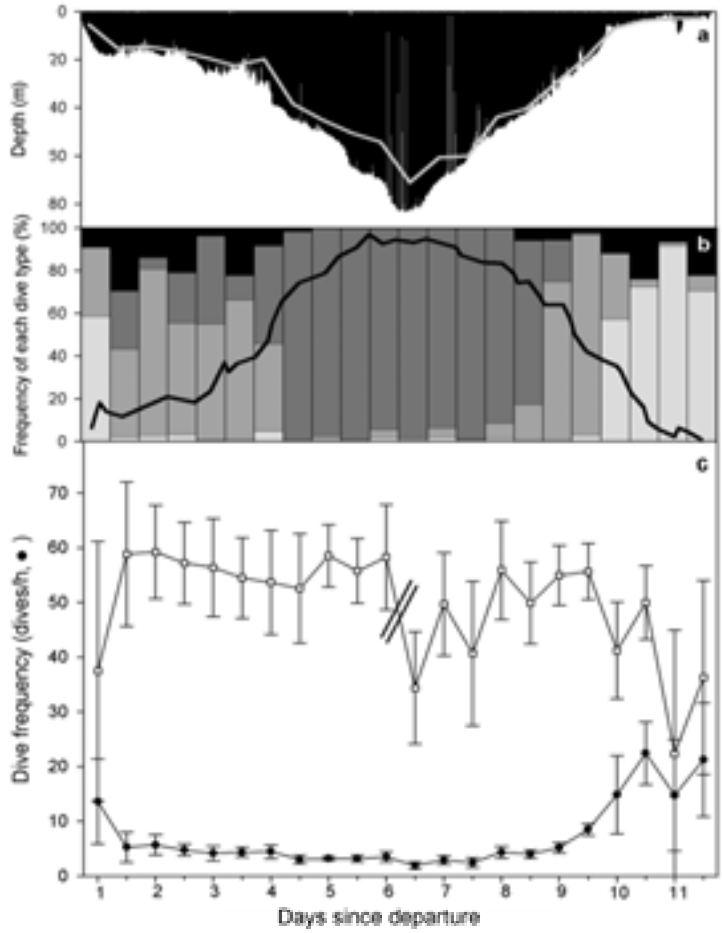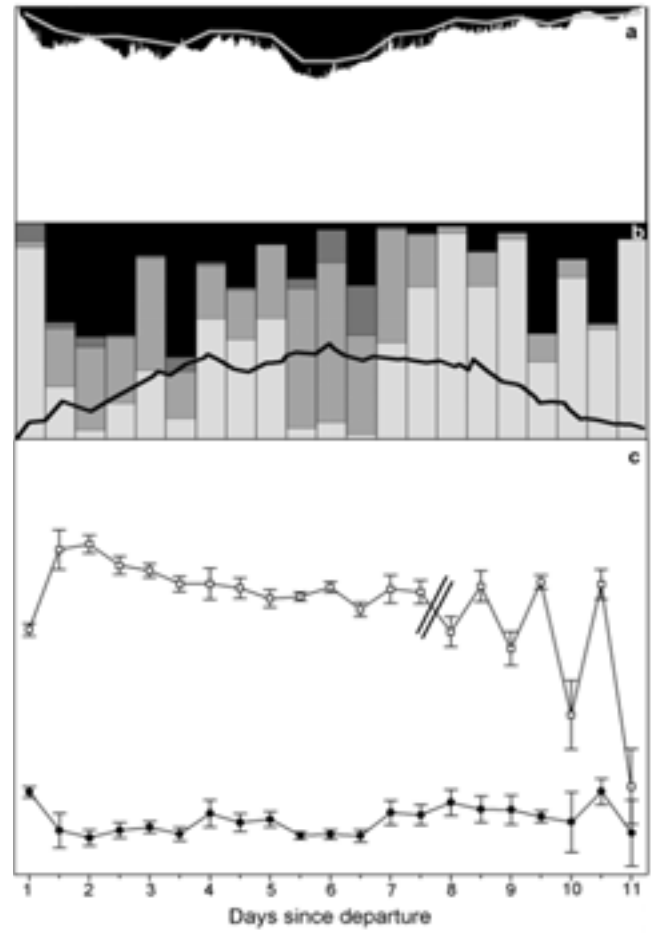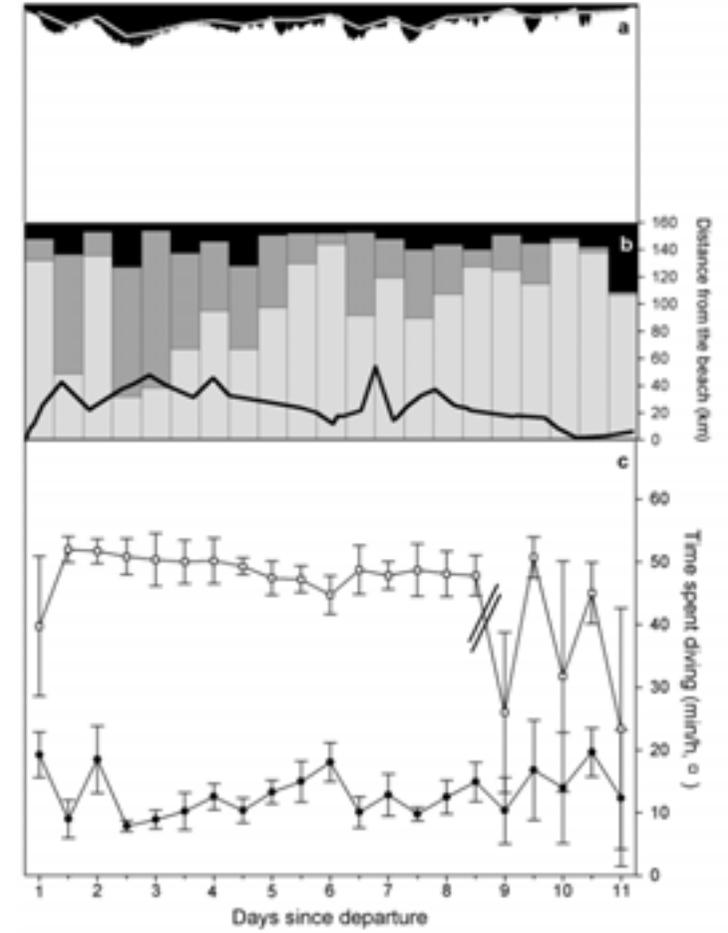

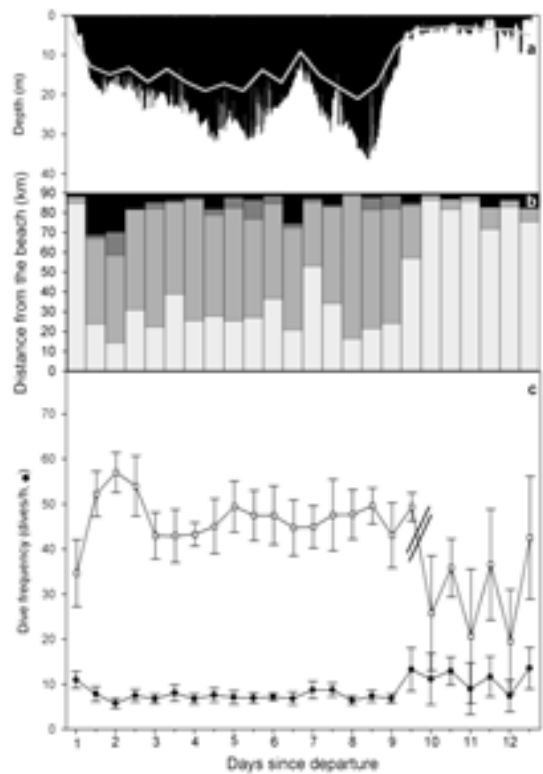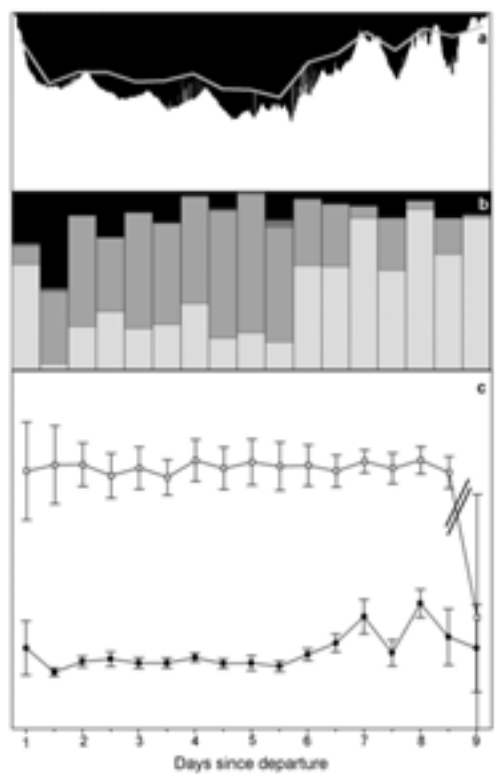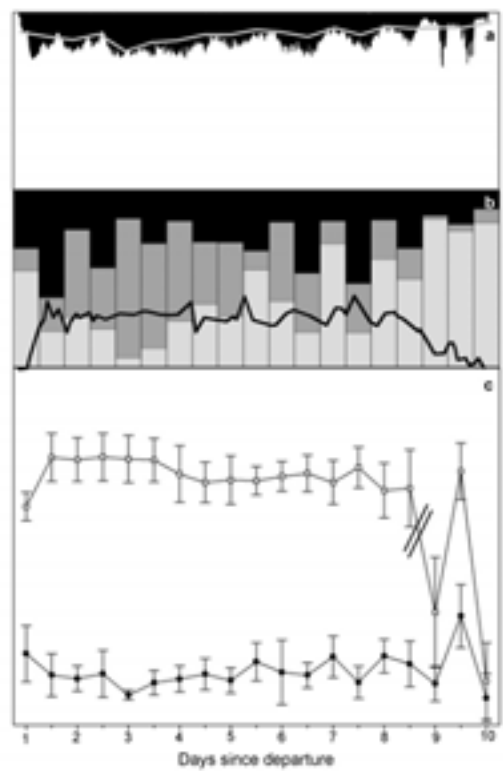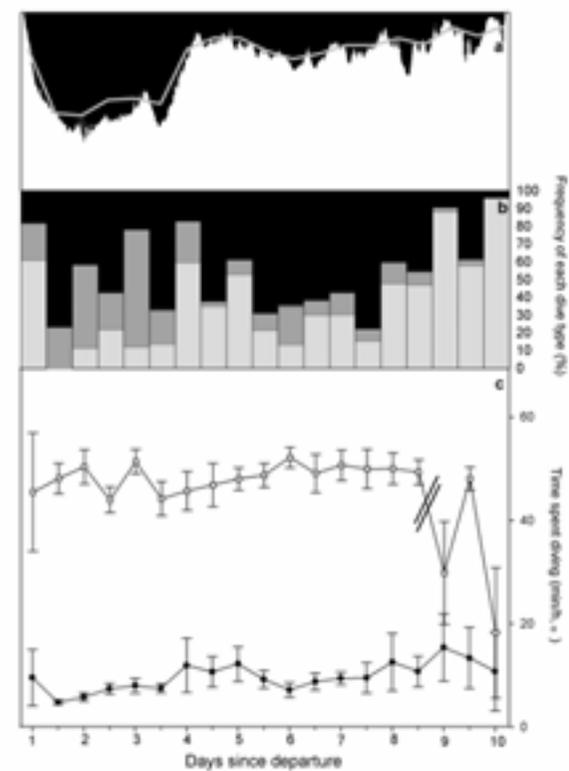